\documentclass[iop]{emulateapj}
 \usepackage{psfrag}
 \usepackage{rotating}
 
\shorttitle{Limb-darkening \& Planetary Transits}
\shortauthors{Neilson~et~al.}

\begin{document}

\title{Limb Darkening and Planetary Transits II: Intensity profile correction factors for a grid of model stellar atmospheres  }

 \author{Hilding R.~Neilson\altaffilmark{1}}
 \altaffiltext{1}{Department of Astronomy \& Astrophysics, University of Toronto, 50 St.~George Street, Toronto, ON, M5S 3H4, Canada}
\email{neilson@astro.utoronto.ca}
\author {John B. Lester\altaffilmark{1,2}}
\altaffiltext{2}{Department of Chemical \& Physical Sciences, University of Toronto Mississauga, Mississauga, ON L5L 1C6, Canada}
\author{Fabien Baron\altaffilmark{3}}
\altaffiltext{3}{Center for High Angular Resolution Astronomy, Department of Physics and Astronomy, Georgia State University, P.O. Box 5060, Atlanta, GA 30302-5060, USA 0000-0002-5074-1128}

\begin{abstract}
The ability to observe extrasolar planets transiting their stars 
has profoundly changed our understanding of these planetary systems.  However,
 these measurements  depend on how well we understand the properties of the host star, such as radius, luminosity and limb darkening.  Traditionally, limb darkening is treated as a parameterization in the analysis, but these simple parameterizations are not accurate representations of actual center-to-limb intensity variations (CLIV) to the precision needed for interpreting these transit observations. This effect leads to systematic errors for the measured planetary radii and corresponding measured spectral features.  We compute synthetic planetary transits using model stellar atmosphere CLIV and corresponding best-fit limb-darkening laws for a grid spherically symmetric model stellar atmospheres.  From these light curves we measure the differences in flux as a function of the star's effective temperature, gravity, mass, and the inclination of the planet's orbit.  
 \end{abstract}

 \section{Introduction} \label{sec:intro}
There are currently more than three thousand confirmed extrasolar planets, many of which were discovered using the \textit{Kepler} space telescope via the transit method. This method has revolutionized our view of planets and the potential for discovering life in the Universe.

Planet transit observations are now so precise that it is possible to characterize the composition and structure of extrasolar planets \citep{Seager2010}.  As more powerful telescopes and satellites become available and surveys are launched, it is expected that in the next decade more Earth-like planets will be discovered that will potentially detect the presence of biomarkers in the atmospheres of extrasolar planets \citep{Rauer2014, Ricker2015}.  

However, even with all of the progress made in the past decade, there remain a number of challenges.  One such challenge is that analyzing planetary transit light curves requires understanding stellar limb darkening, also called the center-to-limb intensity variation (CLIV).  The CLIV is the observed change of intensity from the center to the edge of the stellar disk.  \cite{Mandel2002} developed an analytic model of a planetary transit assuming a simplified parameterization of stellar CLIV, typically as either a quadratic limb-darkening law or a four-parameter law \citep{Claret2000}.  

Representing the CLIV with a simplified limb-darkening law (LDL) has been a reasonable approach for understanding most planetary transit observations, but there are a number of examples where the measured limb darkening disagreed with that predicted from model stellar atmospheres. \cite{Kipping2011a, Kipping2011b} found that limb-darkening parameterizations measured for a sample of \emph{Kepler} transit observations were inconsistent with predictions, raising questions about the physics of stellar atmospheres along with our understanding of planetary transits. \cite{Howarth2011} argued that the differences were the  result of the planet's orbit being inclined relative to our line of sight.  In that case, the measured limb-darkening parameters differed because the transit observations probed only part of the CLIV whereas the LDLs from model stellar atmospheres are constructed from the entire CLIV.  \cite{Howarth2011} was able to resolve those errors for some stars of that sample by fitting limb-darkening coefficients over only part of the CLIV.

In addition to the degeneracy created by the transit inclination,
the representation of the CLIV also impacts attempts to extract 
information about the transiting planet's spectrum and composition from 
the lightcurve.   For example, there have been conflicting claims regarding the composition of the atmosphere of GJ~1214 from transit spectral observations \citep{Croll2011}.  Using near-infrared transit spectra, \cite{Croll2011, Gillon2014} and \cite{Caceres2014} determined that the planet's atmosphere must have a small mean-molecular weight, but that result is contested by other observations \citep{Bean2011, Berta2012}.

Similarly, \cite{Hirano2016, Fukui2016} and others report precisions of the order of 1\% for measuring $R_{\rm{p}}/R_\ast$ for planets orbiting F-type stars. \cite{Almenara2015} reported precisions better than 1\% for planets orbiting an evolved metal-poor F-star.  These results are very precise yet depend on their assumptions of stellar limb darkening.  As such, can we be sure these measurements are accurate?
 
It is becoming increasingly apparent that the current two-, three- and four-parameter limb-darkening laws are simply inadequate for high precision planetary transit models. We showed  \citep[][hereafter Paper 1]{Neilson2016a} that synthetic planetary transit light curves computed directly from model stellar atmosphere CLIV differ from light curves computed from best-fit limb-darkening laws for a solar-like star, where the only difference is the shape of the intensity profile employed. This shows that fitting errors in  planetary transit observations do not come only from errors in the limb-darkening parameters but also from the assumption of a specific type of limb-darkening law.  These errors range from about 100 to a few hundred parts-per-million and vary as a function of wavelength. Similar results were found  independently by \cite{Morello2017}. Hence, assuming a simple limb-darkening law contaminates measurements of extrasolar planet spectra, oblateness and other phenomena.

Limb-darkening laws are not accurate representations of model stellar atmosphere CLIV, particularly near the limb of the star. \cite{Neilson2011} found that, for spherically symmetric model stellar atmospheres, currently favored quadratic limb-darkening laws fit the model CLIV poorly. The \cite{Claret2000} four-parameter law provides a more precise fit, but it is still of limited accuracy near the limb of the star.  This result was confirmed for giant and supergiant stars ($\log g \le 3$) \citep{Neilson2013a}  as well as for dwarf stars \citep{Neilson2013b}.  Specifically, these laws fail for two reasons: the first is  the more complex structure of the CLIV that prevents simple limb-darkening laws from fitting the intensity near the limb of the star, and the second being the inability for best-fit limb-darkening laws to accurately reproduce the stellar flux.

These two differences between model CLIV and best-fit limb-darkening laws cause the differences between synthetic planetary transit light curves found in Paper 1.  Because the errors in best-fit limb-darkening are a function of stellar properties, it is likely that the errors introduced by assuming a simple limb-darkening parameterization are also a function of stellar properties.  In this work, we present computed errors as a function of stellar properties and waveband for dwarf stars. These can be applied to planetary transit observations for the purpose of defining the systematic uncertainties of any fit, as well as determining their impact on additional phenomena such as spectral features and oblateness.  In the next section we describe our models and how we measure the differences between synthetic planetary transit light curves computed directly from model CLIV and from limb-darkening laws. In Section~{sec:radius}, we consider the definition of the stellar radius and its impact in our analysis. We present the errors for our model stellar atmosphere grids in Section~\ref{sec:errors} as a function of stellar properties, and we present our results in Sections~\ref{sec:errors_incline} and \ref{sec:correction}.   We discuss the impact of these results in terms of the atmospheric extension of a star, i.e., the size of the atmosphere relative to the stellar radius, in Section~\ref{sec:extension}.

 \section{Model stellar atmospheres}\label{sec:atmosphere}
 Our analysis used the spherically symmetric model stellar 
atmospheres from \cite{Neilson2013b}, which were computed using the 
\textsc{SAtlas} codes \citep{Lester2008}. These models were computed for stellar masses spanning the range from 
$M_\ast = 0.2$ to $1.4~M_\odot$ in steps of $\Delta M_\ast = 0.3~M_\odot$, 
effective temperatures $T_{\rm{eff}} = 3500$ to $ 8000~$K in steps of 
100~K and surface gravities $\log g = 4$ to $4.75$ in steps of 0.25 dex.  This is equivalent to a range of luminosities from about 0.01 to $15~L_\odot$ and radii from $0.3$ to $2~R_\odot$.
For each model the stellar CLIV was computed at 329 wavel for one thousand points of $\mu$, where $\mu$ is the cosine of the angle formed by a point on the stellar disk and the disk center. The model atmosphere employed in Paper 1 is part of this grid of models.
 
The model CLIVs, integrated over the $BVRIJK$, {\it Kepler}- and 
{\it CoRot}-wavebands, were used to compute the corresponding best-fit 
limb-darkening coefficients for the quadratic limb-darkening law.  We use these CLIV's, calculated using the methods described in Paper 1, and the corresponding best-fit coefficients to compute synthetic planetary transit light curves using the analytic prescription developed by \cite{Mandel2002} for the small-planet assumption, represented by $\rho$ defined as
\begin{equation} \label{eq:def_rho}
\rho \equiv \frac{R_{\rm{p}}}{R_\ast} \leq 0.1.
\end{equation}

  While the small-planet assumption is not perfect, we have shown that the \emph{difference} between light curves follows the same behavior regardless of planet radius.  Furthermore, all we are truly modeling is the difference between CLIV and limb-darkening as a function of $\mu$. We also note that \cite{Morello2017} found similar results using a different prescription for modeling planetary transits.
 
 Using the synthetic planetary transit light curves computed for each model atmosphere using both the CLIV and limb-darkening coefficients, we compute the average difference and the greatest difference for each waveband and model stellar atmosphere for $\rho = 0.1$.  The computed average difference between light curves acts as a measure of the systematic error of the fit for properties, such as relative planet radius, limb-darkening coefficients, and, potentially, secondary quantities such as planetary oblateness and star spots.
 
 The computed flux differences are functions of $\rho$, defined in Equation~\ref{eq:def_rho}. To first order the difference can be written as
 \begin{equation}\label{eq:diff1}
 \Delta f = (I_{\rm{CLIV}} - I_{\rm{LDL}}) \times \rho^2,
 \end{equation}
 where $I_{\rm{CLIV}}$ and  $I_{\rm{LDL}}$ are the intensities from the CLIV and limb-darkening law, respectively. Because of the definition of $\rho$, the average difference scales as the surface area of the planet relative to the star.  For example, if one measures $\rho = 0.05 $  and our model assumes $\rho = 0.1$, then the measured average error will be $(0.05/0.1)^2 = 0.25 \times$ the difference measured in this paper for the same stellar properties. 
  We also compute the root-mean-square (RMS) flux error as a measure of how well the assumption of a quadratic limb-darkening law fits our more realistic CLIV planetary transit light curve.  
  
 
 \section{Definition of the Stellar Radius} \label{sec:radius}
In a model stellar atmosphere there is no ``edge'' that marks the radius of the star and the transition to empty space.  There are several ways to define the stellar radius \citep{Baschek1991}, and we have chosen to use the Rosseland stellar radius, $R_{\rm{Ross}}$, defined as the radius where the Rosseland optical depth, $\tau_{\rm{Ross}}$, has a value of 2/3 because at that radius in the atmosphere the light has $\approx 0.5$ chance of escaping to space without being absorbed.  However, there is still some radiation emitted by the star from above this level, and the structure of  these levels, and the radiation they emit, are different for our spherical models compared to  plane-parallel models.  Also, there are other definitions of the stellar radius that are commonly used.  One is the limb-darkened radius, $R_{\rm{LD}}$, derived from where the disk visibility observed using optical interferometry goes to zero \citep{Wittkowski2004}, though it should be noted that interferometric visibilities are unreliable for visibilities less than $10^{-4}$ \citep{Baron2014}.  To be clear, $R_{\rm{LD}}$ is greater than $R_{\rm{Ross}}$.  In the analysis to follow, we will show that the exact definition of $R$ is inconsequential because we are comparing results found using the CLIV  directly with results using a LDL representation of the same CLIV, and the definition of $R$ essentially cancels out.

In the next section we explore how the representations of the CLIV differ as a function of stellar properties and inclination. 
As in Paper 1, we define the inclination in terms of $\mu$.   The conventional definition of the orbital inclination angle, $i$, is the angle between the orbit plane and the plane of the sky, so that $i = 90^\circ$ is an orbit observed edge-on and $i = 0^\circ$ is an orbit observed face-on.

 We define a new orbital inclination parameter 
 \begin{equation}\label{eq:theta0}
      \theta_0 \equiv 90^\circ - i
  \end{equation}
   and scaling 
     \begin{equation}\label{eq:mu0}
     \mu_0 \equiv \frac{a \cos \theta_0}{R_{\rm{LD}}},  
     \end{equation}
     where $a/R_{\rm{LD}}$ is the normalized separation between the star and the planet.  The purpose for these definitions is to allow a more direct connection between light curves as a function of inclination with CLIV and limb-darkening laws computed as a function of $\mu$. 
   
    With the definition of $\rho$ given in Equation~\ref{eq:def_rho}, we need to return to the definition of the star's radius.  In particular, how do we use the spherically symmetric model stellar atmosphere CLIV to fit planetary transit observations and measure the planet radius itself? We suggest two possibilities and reject a third.

The first possible solution follows if one uses the spherical model CLIV, or uses a limb-darkening law derived from fitting the spherical model CLIV.  In either case, the approach is to fit the observations and then multiply the measured value of $\rho = R_{\rm{p}}/R_{\rm{LD}}$ by the factor         $R_{\rm{LD}}/R_{\rm{Ross}}$ to transform $\rho$ to the Rosseland radius.  Using the CLIV from the models makes $R_{\rm{LD}}/R_{\rm{Ross}}$ readily available.

The second option is to construct a planetary transit code that forces the edge of the stellar disk to be $R_{\rm{Ross}}$ such that $\mu = 0$ corresponds to the point $R_{\rm{LD}}$. However, this method also requires knowing the ratio between $R_{\rm{Ross}}$ and $R_{\rm{LD}}$, so the first option is preferred as being simpler for computation.
   
The third option, which we reject, is to clip the CLIV so that the contribution to the CLIV from the extended part of the atmosphere is removed, and then to rescale the CLIV so that $\mu = 0$ corresponds to $R_{\rm{Ross}}$ \citep{Claret2003, Espinoza2016, Claret2017}. This clipping can be done by knowing where the values $R_{\rm{Ross}}$ and $R_{\rm{LD}}$ are in the model that will be clipped or by assuming that the point in the CLIV where the derivative of the intensity with respect to $\mu$ is greatest. \cite{Aufdenberg2005} has shown that this is approximately the point corresponding to $R_{\rm{Ross}}$. 

However, we reject this option because it removes information about the stellar atmosphere and its radiation properties.  When we clip the CLIV, we remove information about atmospheric extension and make the CLIV more plane-parallel-like. Furthermore, clipping the CLIV and rescaling the intensity profile will increase the moments of the intensity, in particular the stellar flux.  If the stellar flux is increased in a planetary transit fit then the corresponding value of $\rho$ will be smaller. As such, when one clips the CLIV to get a better fit one creates both an inconsistency in the stellar models and biases the fit to smaller values of $\rho$. 

Regardless of the method used to incorporate spherically symmetric model stellar atmospheres into fits of transit light curves, the results remains the same. One can either use model knowledge of $R_{\rm{Ross}}/R_{\rm{LD}}$ to improve the analysis or one can continue to use geometrically-unrealistic models or models with inconsistent fluxes due to clipping that will bias any analysis.  For the sake of this work, the issue is not of consequence since we will show that the analysis is a \emph{relative} comparison.


\section{Measuring the errors}\label{sec:errors}
\cite{Neilson2013a, Neilson2013b} found that the errors produced by fitting limb-darkening laws to spherically symmetric model stellar atmosphere CLIV varied as a function of atmospheric extension.  The extension can be represented as
\begin{equation}\label{eq:atmos_extension}
 H_{\rm{p}}/R_\ast \propto T_{\rm{eff}}R_\ast/M_\ast = T_{\rm{eff}}/(gR_\ast)= T_{\rm{eff}}/\sqrt{gM_\ast}
 \end{equation}
\citep{Baschek1991, Bessell1991, Neilson2016b}.  This extension, also referred to as the stellar mass index (SMI) by \cite{Neilson2016b}, is important because it indicates how the structure of the CLIV changes near the edge of the stellar disk. Because the errors for fitting limb-darkening grow as a function of this extension, we expect the average difference between synthetic light curves  also to increase as a function of atmospheric extension.  

\begin{figure*}[t]
\begin{center}
\plottwo{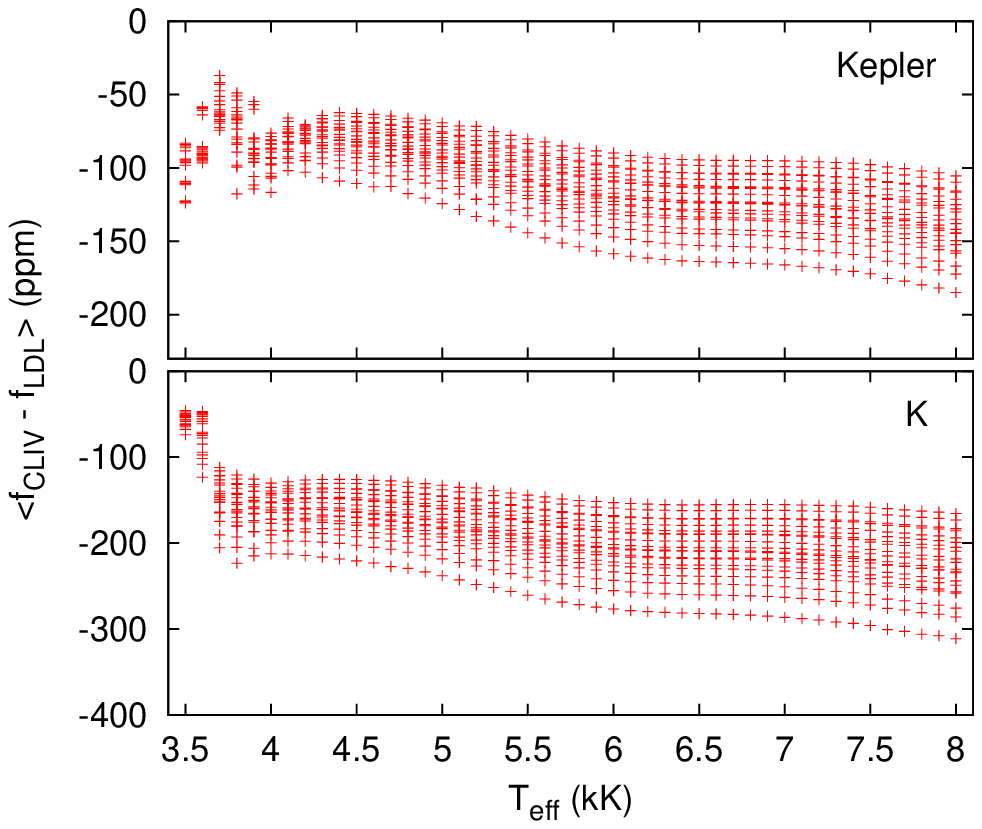}{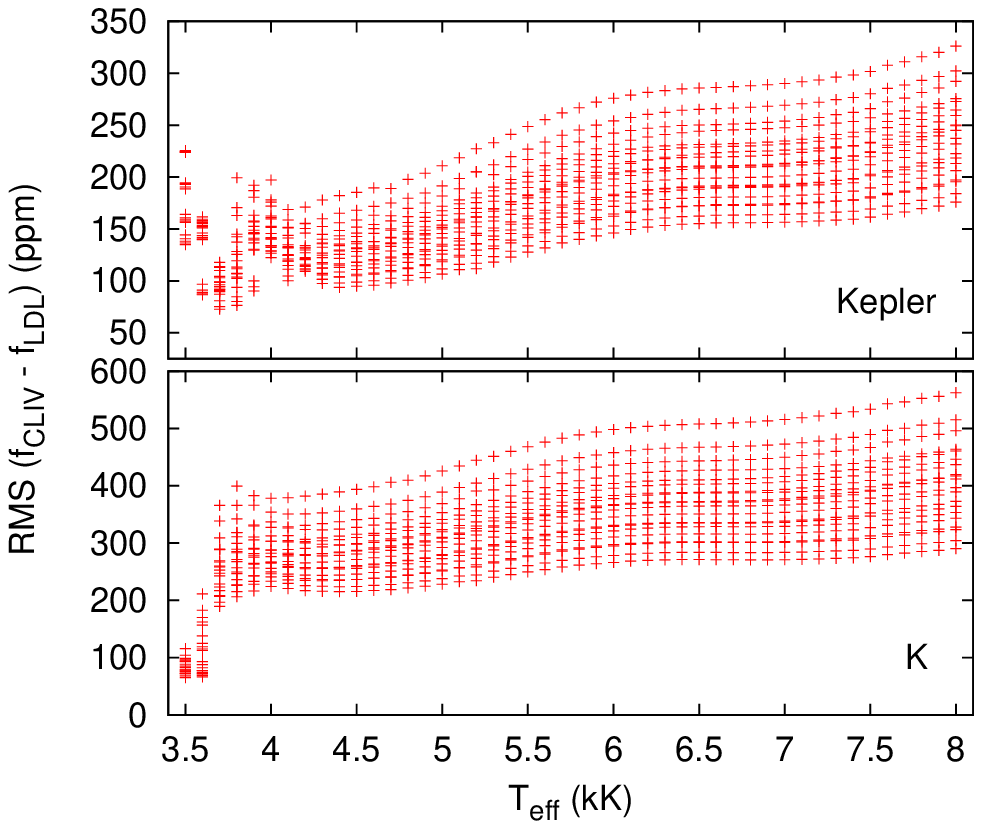}
\end{center}
\caption{(Left) Average differences between synthetic planetary transit light curves computed using model stellar atmosphere CLIV and using best-fit quadratric limb-darkening laws as a function of effective temperature for the {\it Kepler}-band (top) and $K$-band (bottom). (Right) Same as the left panels but for the RMS difference of the light curves.}\label{f1}
\end{figure*}

Before we explore the dependence of the limb darkening on the parameterization of the atmospheric extension, we first consider, for the case of edge-on inclination, $i=90^\circ$, how the average differences change independently as a function of effective temperature, gravity and stellar mass.   Under these assumptions we plot the errors for the {\it Kepler}- and $K$-bands, although we have also computed these differences for $BVRIH$-, and \emph{CoRot}-bands.
In Figure~\ref{f1} we plot the average flux difference between the CLIV and the best-fit quadratic limb-darkening law for an entire transit and the greatest difference during the transit as a function of effective temperature.  It is notable that these differences trend toward greater values with increasing effective temperatures.  Hence,  hotter stars with transiting planets will have greater systematic uncertainties, up to 300~ppm for the {\it Kepler}-band and 600~ppm for the $K$-band. This error in flux, $\Delta f = f_{\rm{CLIV}} - f_{\rm{LDL}}$, is also an error in the surface area of the planet relative to the star, which, for the small planet approximation is $\rho^2=0.01$, hence the errors reach about 3\% and 6\% in the {\it Kepler}- and $K$-bands, respectively. 

\begin{figure*}[t]
\begin{center}
\plottwo{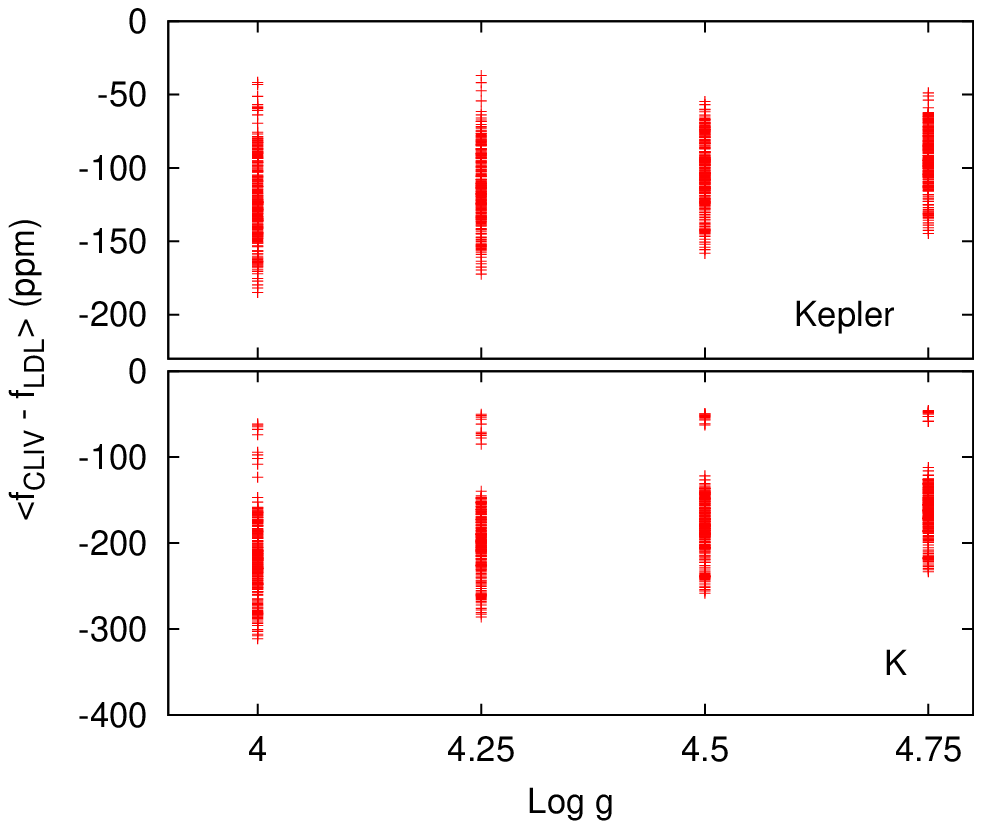}{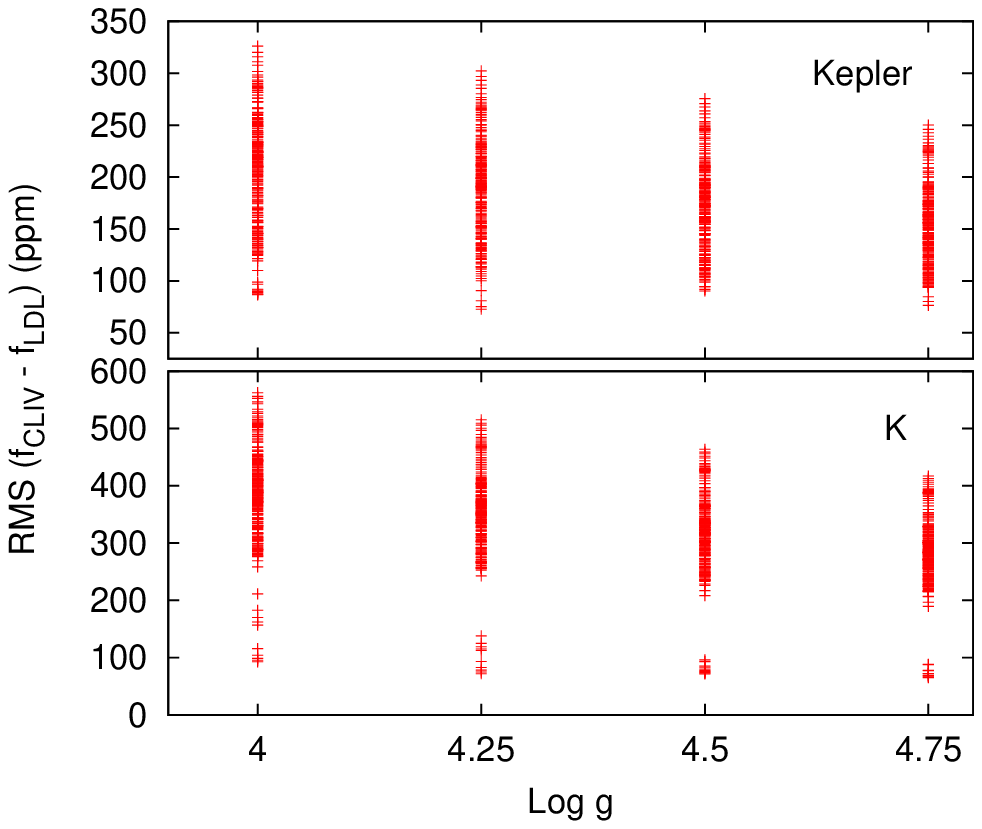}
\end{center}
\caption{(Left) Average differences between synthetic planetary transit light curves computed using model stellar atmosphere CLIV and using best-fit quadratric limb-darkening laws as a function of stellar gravity for the {\it Kepler}-band (top) and $K$-band (bottom). (Right) Same as the left panels but for the RMS difference of the light curves.}\label{f2}
\end{figure*}

\begin{figure*}[t]
\begin{center}
\plottwo{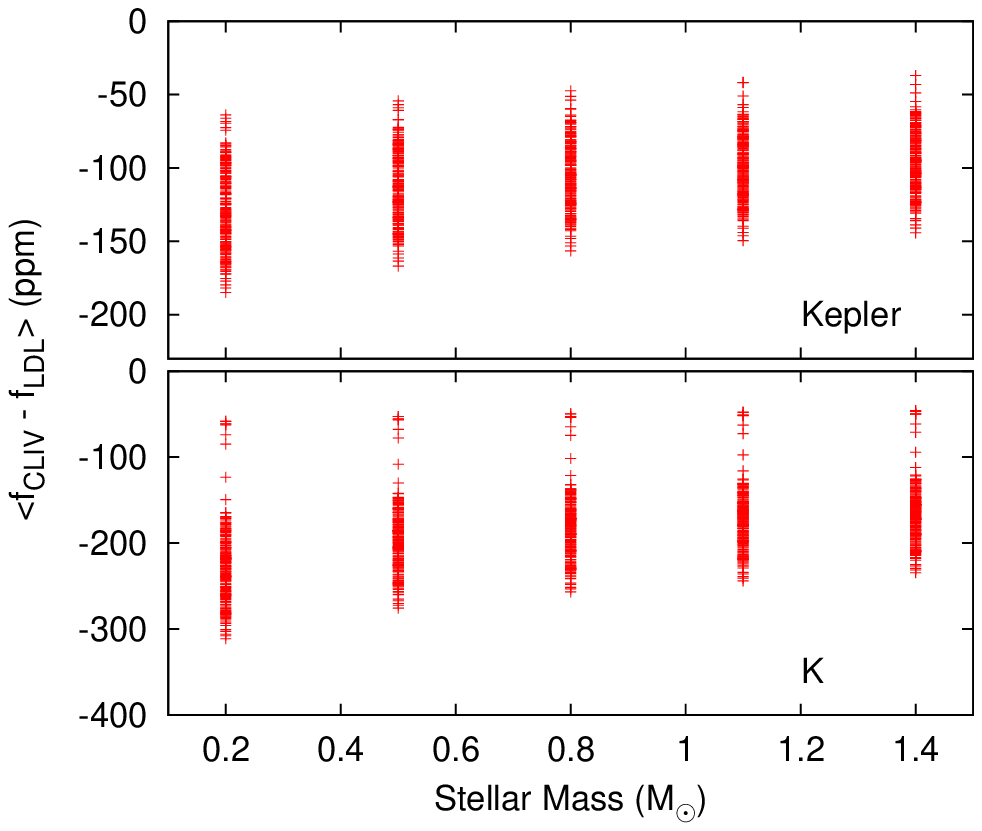}{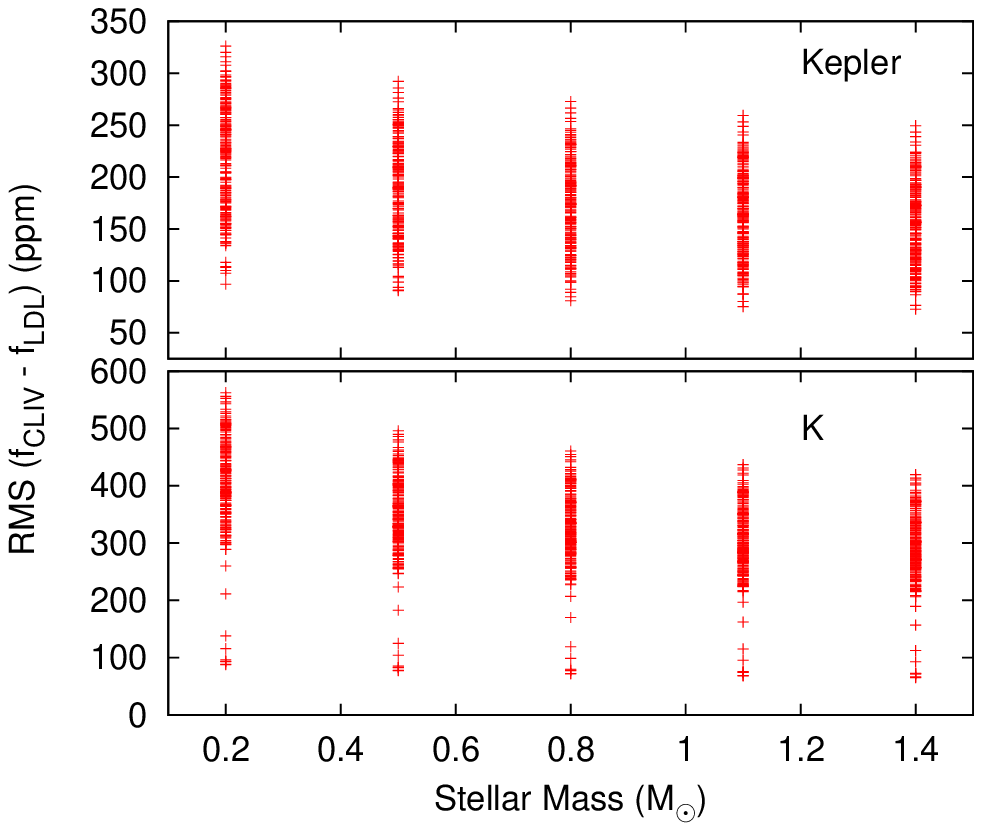}
\end{center}
\caption{(Left) Average differences between synthetic planetary transit light curves computed using model stellar atmosphere CLIV and using best-fit quadratric limb-darkening laws as a function of stellar mass for the {\it Kepler}-band (top) and $K$-band (bottom). (Right) Same as the left panels but for the RMS difference of the light curves.}\label{f3}
\end{figure*}

The errors plotted in Figure~\ref{f1} do have a weak trend with effective temperature, but there is an even more significant spread in the errors, by as much as 200~ppm, at every effective temperature.  Because of this spread, we plot the errors as a function of $\log g$ in Figure~\ref{f2} 
The errors show essentially no dependence on surface gravity,
with just a very slight increase for lower gravity model atmospheres.
This weak dependence on gravity is disappointing because the 
gravity-jitter relation \citep{Bastien2013, Bastien2014} would provide a 
quick and simple connection to the errors if they were more sensitive to 
the surface gravity.  Figure~\ref{f3} plots the errors as a function of stellar mass, which is a component of the surface gravity, showing that there is more of a trend, with the greatest differences occur for the smallest stellar masses.

\begin{figure*}[t]
\begin{center}
\plottwo{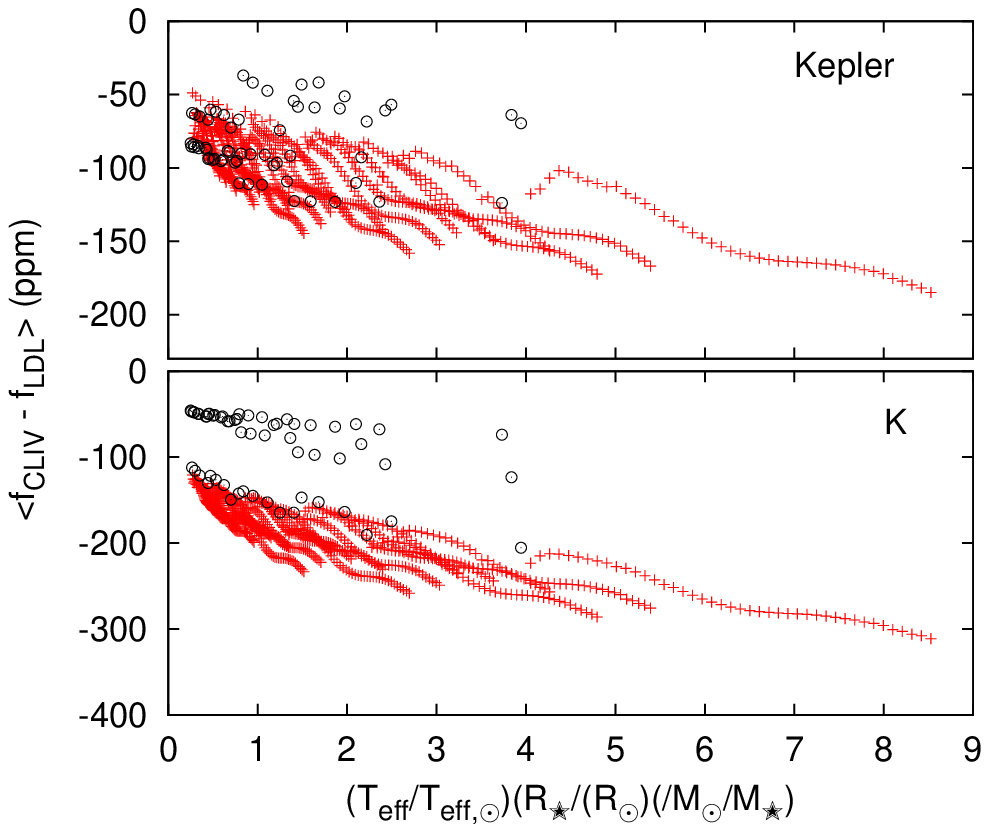}{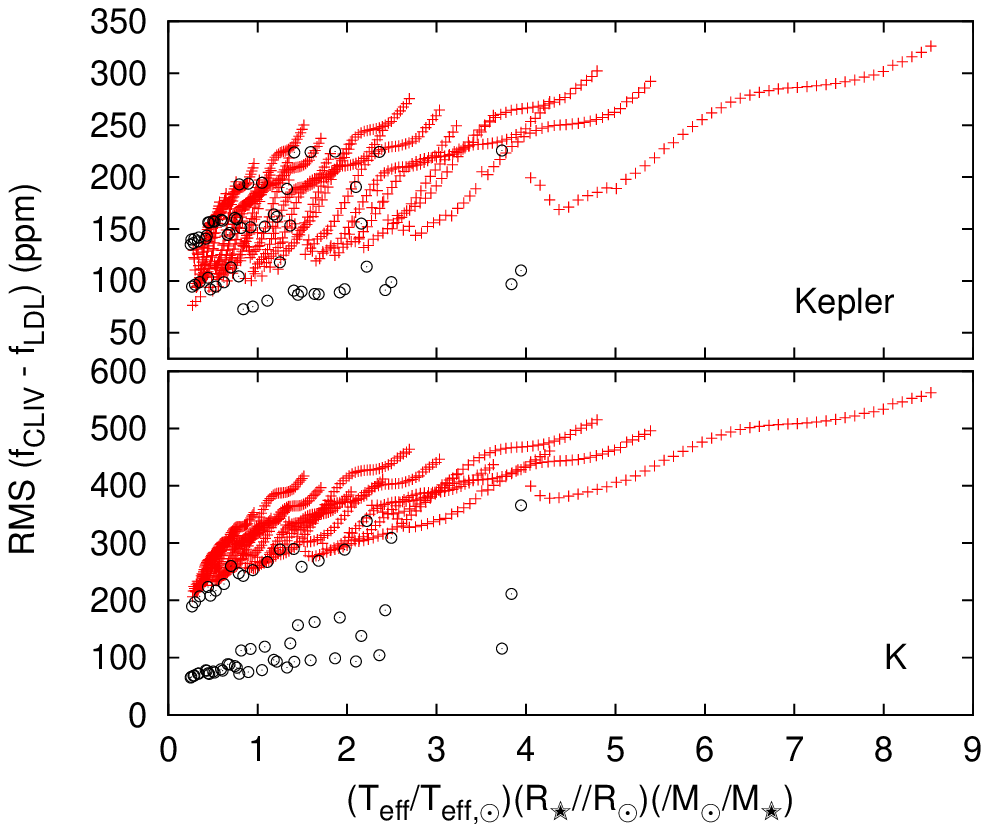}
\end{center}
\caption{(Left) Average differences between synthetic planetary transit light curves computed using model stellar atmosphere CLIV and using best-fit quadratric limb-darkening laws as a function of atmospheric extension for the {\it Kepler}-band (top) and $K$-band (bottom). (Right) Same as the left panels but RMS difference of the light curves. Points denoted by black circles are for model stellar atmospheres with $T_{\rm{eff}} \le 3700$~K.}\label{f4}
\end{figure*}

The results of the three plots imply that the predicted errors trend toward greater absolute values for hotter effective temperature, smaller masses and potentially depends on the stellar gravity.  To test this we use the definition of atmospheric extension given in Equation~\ref{eq:atmos_extension} expressed in solar units.  In Figure~\ref{f4}, we plot the errors versus the atmospheric extension and find there is a trend, though the range of atmospheric extensions is relatively small for these dwarf stars.  \cite{Neilson2016b} computed atmospheric extensions for red giant and supergiant model stellar atmospheres that reach a few hundred $R_\odot/M_\odot$.   In Figure~\ref{f4}, there appear to be two trends: one group that has larger variability and contains most of the models in the sample and a second group with few models and errors that are smallest.  That latter group corresponds to effective temperatures $\le 3700$~K, which likely corresponds to a shift in the dominant opacities in the model stellar atmospheres. 

The key result from Figure~\ref{f4} is that the errors between the atmosphere's actual CLIV and the limb-darkening law representation of this CLIV grows as a function of atmospheric extension.  This result is consistent with the predictions of  \cite{Neilson2013b} that the best-fit limb-darkening coefficients fit the CLIV of model atmosphere most poorly when the models have the greatest extension. The difference between planetary transit light curves computed using model CLIV and those computed using best-fit limb-darkening coefficients is tracing the quality of the fit of those best-fit limb-darkening coefficients.  

The greatest differences correspond to the greatest atmospheric extensions, hence the hottest and most evolved stars in our sample with $T_{\rm{eff}} \rightarrow 8000$~K and $\log g \rightarrow 4.0$. That is, the greatest differences  correspond to evolved main sequence F-type stars. There have been numerous planet transit detection around F-type stars \citep{Gandolfi2012, Smalley2012, Bayliss2013, Huang2015, Fukui2016} and many of those exoplanets appear to be `bloated' hot Jupiters.  Understanding the errors introduced by assuming simple limb-darkening laws could resolve some of this `bloating', especially since we found in Paper 1 that those differences increase when we consider orbits that are inclined from edge on.


\section{The errors as a function of orbital inclination} \label{sec:errors_incline}
In this section, we explore how the differences between synthetic planetary transit light curves computed using model stellar atmosphere CLIV and those computed using best-fit limb-darkening coefficients change as a function of orbital inclination.  We represent the inclination using $\mu_0$, defined in Equation~\ref{eq:mu0}, with an edge-on orbit having $\mu_0 = 1$ and a face-on orbit having $\mu_0 = 0$.   In Paper 1, we found that the differences between light curves can increase with increasing inclination until $\mu_0 \approx 0.3$, which corresponds to  $\theta_0 \approx 70^\circ$, $i \approx 20^\circ$ and impact parameter $b \approx 0.95$, $b \equiv (a/R_\ast)\cos i$, where $a/R_\ast$ is the orbital separation relative to the radius of the star.  As a result, for most orbits a change in inclination will lead to greater errors, and the maximum differences between light curves depend on the inclination even more.

\begin{figure*}[t]
\begin{center}
\plottwo{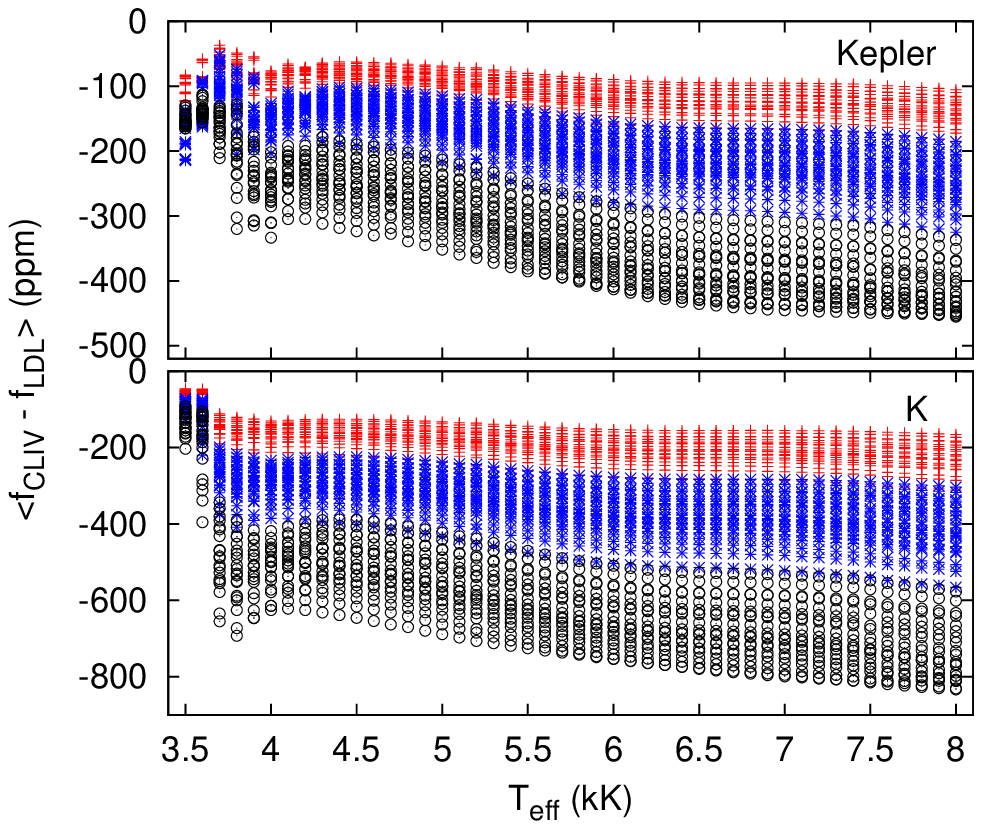}{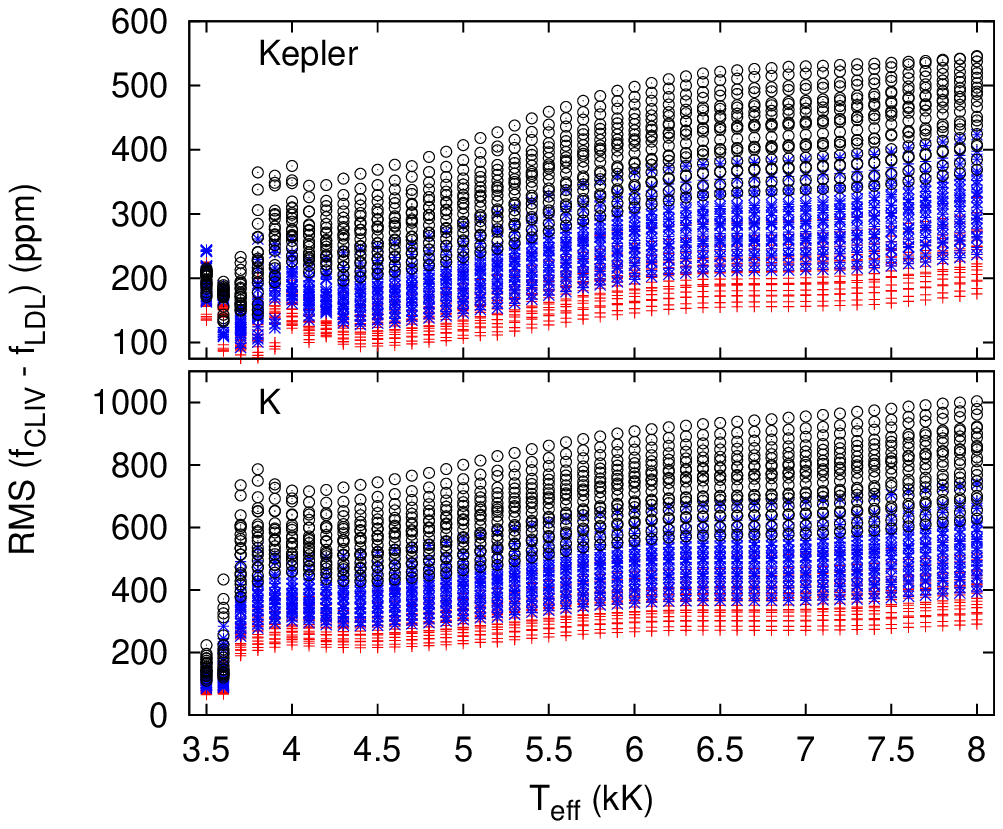}
\end{center}
\caption{(Left) Average differences between synthetic planetary transit light curves computed using model stellar atmosphere CLIV or using best-fit quadratric limb-darkening laws as a function of effective temperature for the {\it Kepler}-band (top) and $K$-band (bottom). (Right) Same as the left panels but for the RMS of the light curves. The red crosses represent transits with $\mu_0 = 1$, blue stars $\mu_0 = 0.7$ and black open squares $\mu_0 = 0.3$. The spread of the average differences and RMS values for each $\mu_0$ arises from variations in stellar mass and gravity for a given stellar effective temperature.}\label{f5}
\end{figure*}
\begin{figure*}[t]
\begin{center}
\plottwo{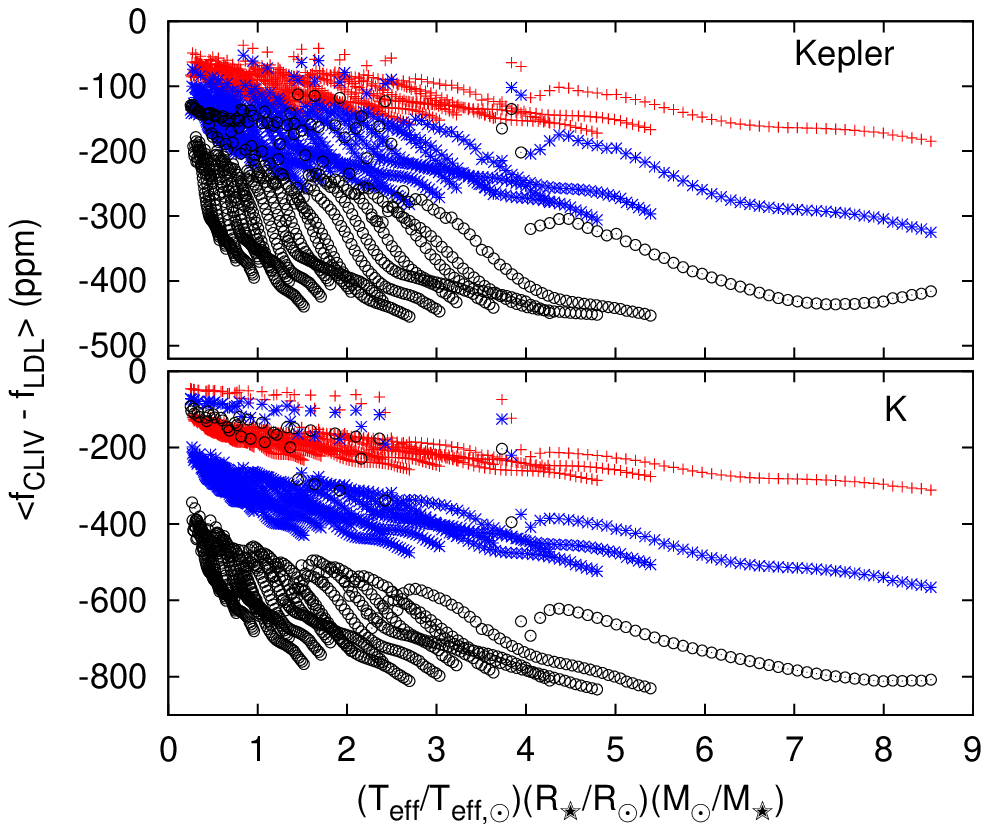}{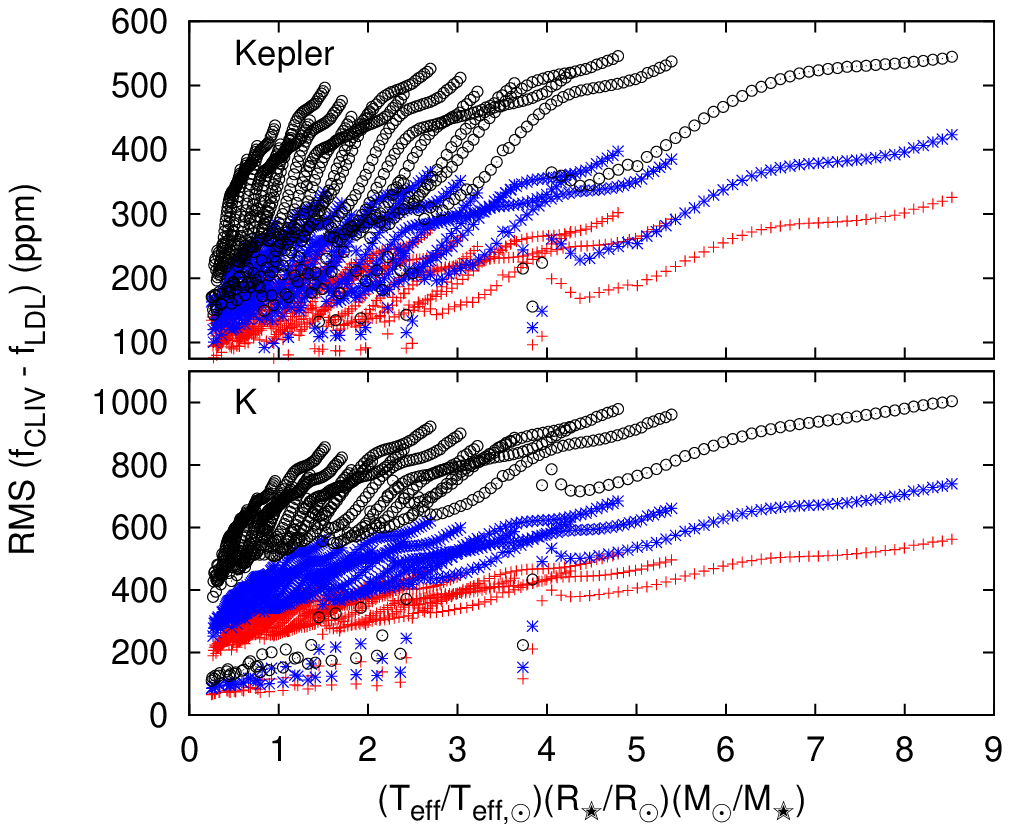}
\end{center}
\caption{(Left) Average differences between synthetic planetary transit light curves computed using model stellar atmosphere CLIV or using best-fit quadratic limb-darkening laws as a function of atmospheric extension for the {\it Kepler}-band (top) and $K$-band (bottom). (Right) Same as the left panels but for the RMS difference of the light curves. The red crosses represent transits with $\mu_0 = 1$, blue stars $\mu_0 = 0.7$ and black open squares $\mu_0 = 0.3$. }\label{f6}
\end{figure*}

We test the role of orbital inclination by first plotting the errors as a function of effective temperature in Figure~\ref{f5}. The errors due to assuming limb-darkening laws increase as a function of orbital inclination.  In the {\it Kepler}-band, the average difference between light curves shifts from about -300~ppm for $\mu_0 = 1$, up to -400~ppm for $\mu_0 = 0.7$ and up to -600~ppm  for $\mu_0 = 0.3$.  In the $K$-band, the effect is even more significant with differences up to -700~ppm and -1000~ppm  for the hottest stars.  As such, we are seeing how connected and dependent the light curve and the assumption of limb-darkening laws are on each other \citep{Howarth2011}.

The maximum differences also grow as a function of orbital inclination. As $\mu_0$ decreases from unity to zero, the maximum difference reaches almost 1600~ppm and 2600~ppm in the {\it Kepler}- and $K$-bands, respectively.  Those differences correspond to about 16\% and 26\% of the surface area of the assumed planet, hence about 8\% and 13\% of the planet radius for $\delta A = 2\delta R_{\rm{p}}$.   For more inclined orbits, we are finding errors that are a significant fraction of the relative planet size.  
 
In Figure~\ref{f6} we show the effect of atmospheric extension 
on the difference between the CLIV and the quadratic limb-darkening law.
The results are surprising.  In Figure~\ref{f4} we found that the average and maximum difference between synthetic planet transit light curves grow as a function of atmospheric extension.  However, we see that for more inclined orbits the average differences increase rapidly as a function of atmospheric extension.  When $\mu_0 = 0.3$ we find that the average difference reaches almost $-500$~ppm and $-1100$~ppm in the {\it Kepler}- and $K$-bands, respectively, with an atmospheric extension of  $\approx 2~R_\odot/M_\odot$.   This suggests that all stars with planets orbiting in inclined orbits will have significant errors for even smaller atmospheric extensions.

These results offer distinct challenges for our understanding of planet transits and secondary effects, such as oblateness, rotation and spots. For instance, for the case of KIC~8462852 \citep{Boyajian2016}  the transits have been explained by large families of orbiting comets (or dust clouds)\citep{Bodman2016}. Because that analysis ignores limb darkening in fitting the family of comets, if any of the orbits are inclined then the sizes required for the comets will be significantly wrong. As such, our results  show that we must treat limb darkening in planetary transits with greater care and should move from assuming the simple parameterizations to using more realistic models of CLIV.

\section{Correcting the planetary radius}\label{sec:correction}
While the average flux difference offers one measure of the error created by assuming a simple limb-darkening law, it does not offer a significant measurement of biases in the predicted planetary radius.  
To address this, we start by defining the $\chi^2$ from the transit model as
\begin{equation}\label{chi1}
\chi^2 \equiv \sum_z \left[ f_{\rm{CLIV}}(\rho, z) - f_{\rm{LDL}}(\rho, z)\right]^2.
\end{equation}
 In Equation~\ref{chi1}, $z$ is  the projected separation between the center of the planet and the center of the star normalized by the stellar radius. At the edge of the stellar disk $z = 1$.  This definition offers potential challenges for working with CLIV computed for stellar models with atmospheric extension that we will discuss in Sect.~\ref{sec:extension}.
We note this is assuming the small-planet approximation for ease, which will differ slightly from more exact methods.  However, this analysis will allow us to probe the order-of-magnitude of the predicted errors.
Because both the CLIV and the best-fit limb-darkening coefficients
use the same planet radius and inclination, this $\chi^2$ should ideally be a minimum.  However, it is possible to gain improvement by varying some of the parameters.  For instance, varying the limb-darkening coefficients changes the predicted stellar flux, which will change the transit depth and,  hence, will lead to a different value of the planetary radius.  However, this change in limb-darkening coefficients will compound errors in how we understand the host star. Similarly, varying the inclination creates biases in the measured limb-darkening coefficients that alter a fit in the same direction.  For simplicity, we minimize the $\chi^2$ function using just the variation of the planetary radius.  
\begin{figure*}[t]
\begin{center}
\plottwo{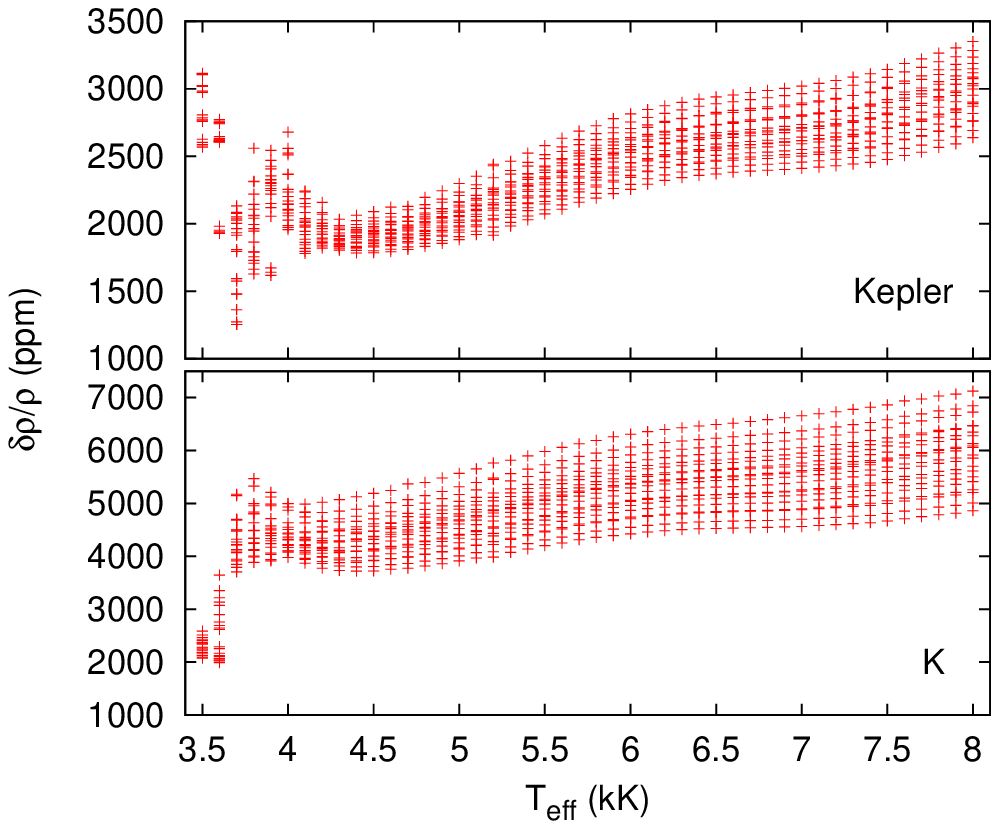}{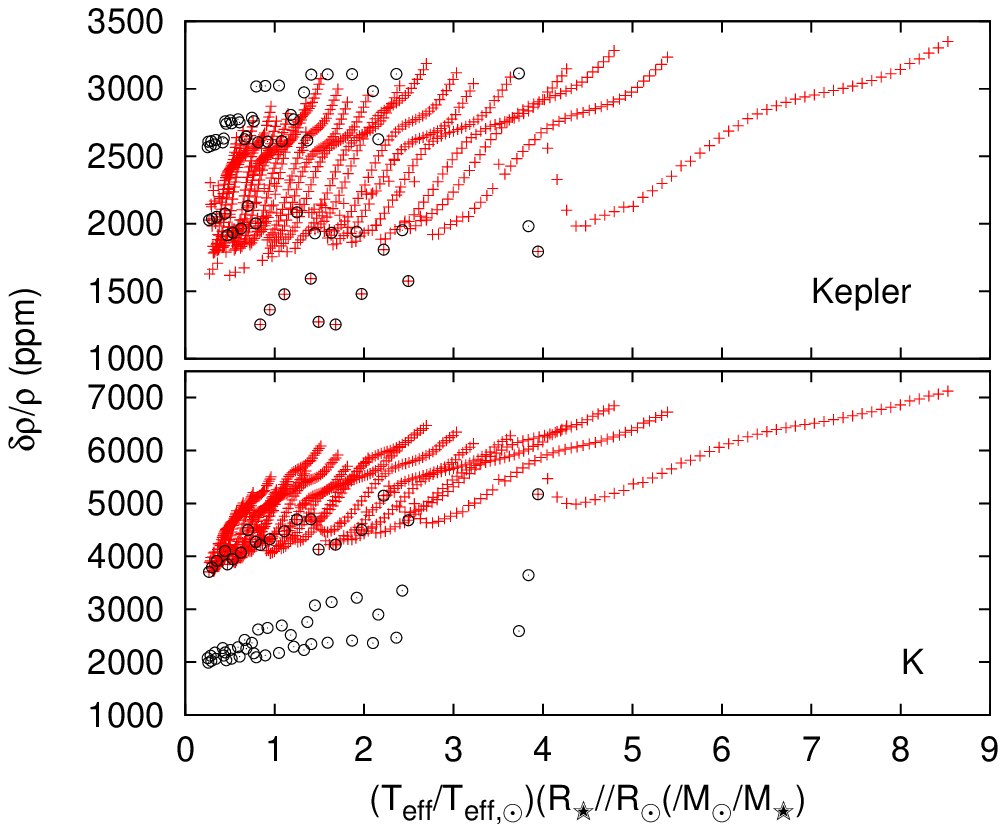}
\end{center}
\caption{Predicted overestimated value of the planetary radius relative to the actual planet size as a function of (Left) effective temperature and (Right) atmospheric extension for the {\it Kepler}- and $K$-bands.  Points denoted by black circles in the right plot are for model stellar atmospheres with $T_{\rm{eff}} \le 3700$~K. The difference shown here is solely caused by assuming a quadratic limb-darkening law. }\label{fig:drho}
\end{figure*}

We start by perturbing the radius in the LDL light curve of Equation~\ref{chi1}
\begin{equation}\label{chi2}
\chi^2 \propto \sum_z \left[ f_{\rm{CLIV}}(\rho, z) - f_{\rm{LDL}}(\rho + \delta \rho, z)\right]^2.
\end{equation}
Next we assume the small-planet approximation, 
\begin{equation}\label{eq:flux0}
f(\rho,z) = 1 - \rho^2 \frac{I^*}{4\Omega},
\end{equation}
is valid for both the CLIV and LDL light curves.  In 
Equation~\ref{eq:flux0} $4\Omega$ is the stellar flux and $I^*$ is the amount of flux blocked by the planet as it transits \citep{Mandel2002}.  For the purpose of this perturbation, we ignore changes in $I^*$ as a function of planet radius as well as second-order changes in 
$\delta \rho$. Therefore

\begin{eqnarray}\label{eq:fldl_drho}
f_\mathrm{LDL}(\rho + \delta \rho, z) & = & 1 - \rho^2 \frac{I^*}{4\Omega} 
                    - 2 \frac{\delta \rho}{\rho} \rho^2  \frac{I^*}{4\Omega} \nonumber \\
& = & f_{\rm{LDL}}(\rho, z) - 2 \frac{\delta \rho}{\rho}\left[1 - f_{\rm{LDL}}(\rho,z)\right].
\end{eqnarray}


We now minimize the $\chi^2$-function, Equation~\ref{chi2}, with
respect to radius to get
\begin{equation}\label{eq:dchi2}
\frac{d\chi^2}{d\rho} = 2 \sum_z \left[ f_{\rm{CLIV}}(\rho, z) - f_{\rm{LDL}}(\rho + \delta \rho, z)\right]\frac{d f_{\rm{LDL}}}{d\rho} = 0.
\end{equation}
Again ignoring changes in $I^*$ as a function of $\rho$, the derivative of Equation~\ref{eq:flux0} gives
\begin{equation}\label{eq:dfldl}
\frac{d f_{\rm{LDL}}}{d\rho} = -2\rho \frac{I^*}{4\Omega} = - \frac{2}{\rho}\left[1 - f_{\rm{LDL}}(\rho,z)\right].
\end{equation}
Using Equations~\ref{eq:fldl_drho} and \ref{eq:dfldl} in 
Equation~\ref{eq:dchi2} gives
\begin{eqnarray*}
&&\sum_z \left\{\left[ f_{\rm{CLIV}}(\rho, z) - f_{\rm{LDL}}(\rho, z) + 2\frac{\delta \rho}{\rho}\left(1 - f_{\rm{LDL}}(\rho,z)\right)\right] \right. \\ 
&&\left. \left[-\frac{2}{\rho}\left(1 - f_{\rm{LDL}}(\rho,z)\right)\right]\right\} = 0.
\end{eqnarray*}
Rearranging and solving for $\delta \rho/\rho$ leads to
\begin{equation}\label{eq:drho}
\frac{\delta \rho}{\rho} = \frac{ \sum_z \left[ (f_{\rm{LDL}} - f_{\rm{CLIV}})(1 - f_{\rm{LDL}}) \right]}{2\sum_z (1- f_{\rm{LDL}})^2}.
\end{equation}
This shows again that the average difference in flux offers a rough measure of the error of the fit that affects the predicted depth of the transit and hence the measured planet radius.   

We note that Equation~\ref{eq:drho} appears to be an explicit function of $\rho$ since $f_{\rm{LDL}}$ and $f_{\rm{CLIV}}$ themselves also depend on $\rho$.  However, if we insert Equation~\ref{eq:flux0} into Equation~\ref{eq:drho} the relative radius cancels leaving only terms of $I^*/4\Omega$.  The amount of flux blocked by the planet, however, is implicitly dependent on the size of the planet, but to first order Equation~\ref{eq:drho} is independent of planet radius.  While one could measure these differences using fitting codes, this analysis illustrates how the  relative planet radius depends on understanding the limb darkening.  Furthermore, we note that this relation implies that the result is independent of stellar radius in that we can replace $\delta \rho/\rho = \delta r_p/r_p$.

 Figure~\ref{fig:drho} shows the relative correction to the planet's radius due to assuming the quadratic limb-darkening law.  This correction is the expected overestimation of the planet radius by fitting methods that assume this limb-darkening law. We plot the difference of the planet's radius as a function of effective temperature and atmospheric extension.  These differences scale with approximately the same behavior as the plots of RMS($f_{\rm{CLIV}} - f_{\rm{LDL}}$). Furthermore, the differences in the planet's radius are significantly greater than indicated by the average flux difference by almost a factor of twenty in the {\it Kepler}-band and by more than a factor of twenty in the $K$-band. We see again that the correction to the planet's radius is also greatest for stars with the greatest effective temperature and atmospheric extensions.

As a result, we find that  planetary radii can be overestimated by up to 7000~ppm in the near-IR and 3500~ppm in the optical {\it Kepler}-band.  This overestimation is small relative to the assumed size of the planet, especially when the correction is relative to the measured planet size.  Assuming the small planet approximation, $\rho = 0.1$, then $\delta \rho = 700$~ppm and $350~$ppm in the optical and near-IR, respectively.

\begin{figure}[t]
\begin{center}
\plotone{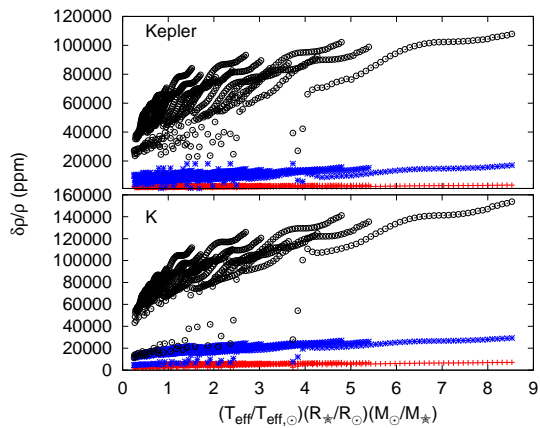}
\end{center}
\caption{Predicted overestimated value of planetary radius relative to the actual planet size as a function of atmospheric extension for different inclinations in the {\it Kepler}- and $K$-bands. The red crosses represent transits with $\mu_0 = 1$, blue stars $\mu_0 = 0.7$ and black open squares $\mu_0 = 0.3$. }\label{fig:drho-b}
\end{figure}

The correction for the planetary radius increases with the inclination of the orbit, similar to that seen for the average flux difference. For instance, Figure~\ref{fig:drho-b} shows that the correction increases by an order of magnitude as $\mu_0 \rightarrow 0$. For the most inclined orbit, and assuming the best-fit limb-darkening coefficients for a quadratic limb-darkening law for each model, the radius correction is about 10\% of the actual planet radius for model stellar atmospheres with the greatest atmospheric extension.  This correction will be even greater if we assume even simpler limb-darkening laws, such as a linear law, or a uniform-disk model (\emph{i.e.}, no limb darkening). Therefore, care must be taken to choose the appropriate limb-darkening parameterization or model when measuring precision values of extrasolar planet radius.

Just for interest, we also computed the error in the planetary radius if one assumes a star with no limb darkening, i.e., a uniform-disk model.  If one uses this method then the planet radius will be underestimated instead of overestimated by about 2 - 5\%.  This check is consistent with the need for more and more free parameters to  precisely measure limb darkening and its impact on the exoplanet radius.  However, our analysis and past works highlights the fact that the stellar CLIV cannot be accurately represented by simple functions \citep{Neilson2011, Neilson2013a, Neilson2013b}.

\section{Atmospheric extension, stellar radius and transits}\label{sec:extension}

The issue of understanding the stellar radius is important not just for planetary transit fits using spherical model atmospheres, but also for fits using plane-parallel models and limb-darkening laws.  The geometry of plane-parallel models contains no information about the stellar radius. As such, it is merely assumed that measurements of stellar radii from asteroseismology or stellar evolution models correspond with the stellar radii in planetary transit fits. Similarly, best-fit limb-darkening laws based on plane-parallel models or fit directly to observations make the same assumption and it is unclear that this is true. Spherically symmetric model stellar atmospheres explicity contain information about the stellar radius, and therefore about the extension of the atmosphere.  Because of the atmospheric extension the edge of the star is at a physical radius that is not the Rosseland radius.  As such plane-parallel and spherical model stellar atmospheres should not be expected to give the same results when fit to planetary transit or interferometric observations because they make different assumptions about the structure of the photosphere.  The challenges for measuring limb darkening and stellar radii (or angular diameters) has been discussed in detail by numerous authors \citep{Wittkowski2004, Neilson2008, Baron2014, Kervella2017}. 

Just as the differences between plane-parallel and spherically symmetric model stellar atmospheres lead to different measurements of stellar radii, they are also fit with different precision by various limb-darkening laws.  \cite{Neilson2013a,Neilson2013b} showed that six different commonly used limb-darkening laws fit plane-parallel models with much better precision than spherically symmetric models with the same fundamental parameters.  The source of this difference is the point of inflection in spherically symmetric CLIV that is a result of including physics of atmospheric extension.  Therefore, current limb-darkening laws do not fit the effects of atmospheric extension.  This result was found by other works such as \cite{Claret2003} and \cite{Espinoza2016}.  However, these works avoid the challenge of fitting atmospheric extension by clipping the spherically symmetric model CLIV to remove all information about the extension.

 \cite{Ligi2016} found uncertainties in measuring angular diameters of exoplanet-host stars to be about 1.9\%.  This is much greater than the atmospheric extension of these stars.  For instance, the Sun has an extension of the order of 0.1\%, based on the ratio of the pressure scale height in the atmosphere and the solar radius.  Therefore, these issues around the definition of stellar radius will not be readily apparent for direct measurements. On the other hand, \cite{Mann2017} measured the relative planet radii for three exoplanets to a precision of $\delta \rho/\rho \approx 2-4\%$, while \cite{Murgas2017} reported precisions of the order 1\% and better.  Furthermore, the next generation of interferometric observations promise to measure angular diameters to about 0.5\% precision \citep{Zhao2011}. At these uncertainties, the biases introduced by assuming the unphysical limb-darkening laws is becoming important, especially as we attempt to measure spectral properties of exoplanets.

\section{Summary}\label{sec:summary}

In this work, we have taken the CLIVs from the \cite{Neilson2013b} grid of spherically symmetric model stellar atmospheres and the corresponding best-fit limb-darkening coefficients for the quadratic limb-darkening law  and computed the differences between synthetic planetary transit light curves using the prescription described by \cite{Neilson2016a}.  
 We evaluated the error resulting from the use of the limb-darkening laws by computing both the average difference and the greatest difference between the CLIV and LDL transit light curves. These differences were computed as a function of fundamental stellar parameters: effective temperature, gravity and stellar mass along with the inclination of the orbit, which we parameterized as $\mu_0 = \cos (90^\circ - i)$.

The results are striking.  Before considering the role of inclination, we found that the average differences between CLIV and limb-darkened transit light curves increased as a function of atmospheric extension, which implies that the average differences are greatest for more evolved F-type stars.  When inclination is included the differences increase significantly  and depend on the atmospheric extension, indicating the errors are roughly similar for most atmospheric extensions.  These negative errors tell us that  the relative planetary radii are being overestimated, especially for the F-stars, by as much as 5\% and at least 1\% for an edge-on orbit in the {\it Kepler}-band.   \cite{Hirano2016, Fukui2016} and others report precisions of the order of 1\% for measuring $R_{\rm{p}}/R_\ast$ for planets orbiting F-type stars. \cite{Almenara2015} reported precisions better than 1\% for planets orbiting an evolved metal-poor F-star.  Given that our models show that $R_{\rm{p}}/R_\ast$ are overestimated, then these measurements have systematic error of at least 1\% that is not accounted for in the fits.

We note that these errors are for the ideal situation where one knows the inclination and where the limb-darkening coefficients are the most accurately determined. Our analysis does not consider the cases where inclination, limb-darkening coefficients and relative radii are fit simultaneously.  In those cases limb-darkening coefficient measurements can deviate significantly from those of model stellar atmospheres \citep{Kipping2011a, Kipping2011b}, implying a strong dependence of the limb darkening on other fitting parameters.  As the limb-darkening coefficients deviates then so too will the errors.  This may not change the errors much, but is something that must  be explored in greater detail.

One key conclusion of our work is that we need to measure stellar CLIV both precisely and directly.  It is becoming clear that our current assumptions of simple limb-darkening laws are just not good enough for understanding the planetary transit observations.  Interferometry is proving to be one method for directly inferring stellar CLIV \citep{Baron2014, Armstrong2016, Kervella2017}.   We recently showed that we can use interferometric measurements in combination with spectroscopy and spherically symmetric model stellar atmospheres to measure stellar fundamental parameters including stellar masses. That result is based on measurements of atmospheric extension in stars.  We suggest that method will be more robust if combined with planetary transit observations as part of a global fit of stellar and planetary parameters.  That work and this is part of an ongoing research project to test limb-darkening and stellar radii measurements from interferometric observations against state-of-the-art model stellar atmospheres.  But, the results of our current work are clearly showing that we are reaching the limits of plane-parallel model CLIV and arbitrary limb-darkening laws that have no physics basis.

From this analysis, we produced corrections of the relative radius of an exoplanet $\delta \rho/\rho$ for a grid of stellar atmosphere models for the wavebands $BVRIHK$ and the {\it Kepler}- and {\it CoRot}-bands that are publicly available.  While it is preferable to fit the model CLIV to transit light curves and to  shift from measuring limb-darkening coefficients to measuring stellar properties, these correction factors can help improve the precision of planetary transit fits of transit spectra.

\acknowledgements{J.B.L. is grateful for funding from NSERC discovery grants. F.B. acknowledges funding from NSF awards AST-1445935 and AST-1616483. H.R.N. and J.B.L. acknowledge that the University of Toronto operates on traditional land of the Huron-Wendat, the Seneca, and most recently, the Mississaugas of the Credit River. The authors are grateful to have the opportunity to work on this land.}
\bibliographystyle{aa}
\bibliography{ldp2}

\end{document}